# Production of nitric oxide by a fragmenting bolide: An exploratory numerical study

Mihai L. Niculescu[1,*], Elizabeth A. Silber[2,3], Reynold E. Silber[4]

[1]INCAS – National Institute for Aerospace Research 'ElieCarafoli', Flow Physics Department, Numerical Simulation Unit, Bucharest 061126, Romania

[2]Department of Earth Sciences, Western University, 1151 Richmond St., London, ON, N6A 3K7, Canada

[3]AstrumPrime Space Research Inc., Dartmouth, NS, B2W6C4, Canada

[4]Department of Geology and Geophysics, Yale University, PO Box 208109, New Haven, CT, 06520-8109, USA

* Corresponding author

**Abstract**

A meteoroid's hypersonic passage through the Earth's atmosphere results in ablational and fragmentational mass loss. Potential shock waves associated with a parent object as well as its fragments can modify the surrounding atmosphere and produce a range of physico-chemical effects. Some of the thermally driven chemical and physical processes induced by meteoroid-fragment generated shock waves, such as nitric oxide (NO) production, are less understood. Any estimates of meteoric NO production depend not only on a quantifiable meteoroid population and a rate of fragmentation, with a size capable of producing high temperature flows, but also on understanding the physical properties of the meteor flows along with their thermal history. We performed an exploratory pilot numerical study using ANSYS Fluent, the CFD code, to investigate the production of NO in the upper atmosphere by small meteoroids (or fragments of meteoroids after they undergo a disruption episode) in the size range from $10^{-2}$ m to 1 m. Our model uses the simulation of a spherical body in the continuum flow at 70 and 80 km altitude to approximate the behaviour of a small meteoroid capable of producing NO. The results presented in this exploratory study are in good agreement with previous studies.

## 1. Introduction

Upon entering the Earth's upper atmosphere, extraterrestrial objects of both cometary and asteroidal origin (sizes ~0.01- to 100 m) [e.g., *Ceplecha et al.*, 1998, *Stulov*, 2006, *Stulov*, 2008] experience strong ablation and fragmentation [e.g., *Baldwin and Sheaffer*, 1971, *Ivanov and Ryzhanskii*, 1997, *Artemieva and Shuvalov*, 2001, *Ceplecha and Revelle*, 2005, *Barri*, 2010, *Gritsevich et al.*, 2011, *Park and Brown*, 2012]. The ablation and fragmentation are the main mechanisms that account for the meteoroid mass loss [*Silber et al.*, 2018b, *Stulov*, 2006] and coincide with the deposition of a large amount of energy in the local atmosphere [e.g., *Register et al.*, 2017, *Wheeler et al.*, 2017]. This rapid energy conversion process also results in heating and a range of subsequent physico-chemical processes in the local atmosphere [e.g., *Silber et al.*, 2018b]. Moreover, the hypersonic centimetre-sized or larger extraterrestrial objects generate strong shock waves [*Silber et al.*, 2018b, *Silber et al.*, 2017], which may influence not only fragmentation [*Artemieva and Shuvalov*, 1996, *Artemieva and Shuvalov*, 2001] but also the rate and type of ablation [*Bronshten*, 1983, *Silber et al.*, 2018b].

In conceptual terms, meteor fragmentation can be defined as the process that involves separation of individual fragments from the hypersonically propagating parent body under the stress of aerodynamic loading. Separated fragments expand rapidly and their trajectory may deviate significantly from the axis of meteoroid propagation. In principle, fragmentation is well understood and it takes place in meteoroids across a range of compositions and velocities [e.g. *Svetsov et al.*, 1995, *Artemieva and Shuvalov*, 2001, *Revelle and Ceplecha*, 2002, *Ceplecha and Revelle*, 2005, *Barri*, 2010, *Gritsevich et al.*, 2011, *Park and Brown*, 2012, *Register et al.*, 2017]. The process of fragmentation also depends on the internal strength and composition of the meteoroid [e.g., *Ceplecha and McCrosky*, 1976, *Cotto-Figueroa et al.*, 2016]. The study of fragmentation is also important toward understanding the parent bodies of meteoroids [e.g., *Popova et al.*, 2011].

There might be several distinct episodes of fragmentation during the meteoroid hypersonic flight in the atmosphere. It is possible to recognize two basic fragmentations regimes: (1) continuous fragmentation; and (2) sudden (or gross) fragmentation [e.g., *Ceplecha et al.*, 1998, *Ceplecha* *and Revelle*, 2005, *Popova and Nemchinov*, 2008]. In terms of numerical modeling of the fragmentation phenomenon, recent hybrid models [e.g., *Register et al.*, 2017, *Wheeler et al.*, 2017] are able to account for discrete fragmentation, and the energy deposition by individual fragments and dust particles. However, both the mathematical and numerical descriptions of the complex multifaceted physical process of fragmentation during hypervelocity atmospheric entry remains one of the challenges in meteor science.

Following a fragmentation episode, fragments may decelerate relative to the initial velocity of the parent body as a function of the angle of the departure from the main body, and complex interaction with the flow field and the shockwave envelope [e.g., *Artemieva and Shuvalov*, 1996, *Artemieva and Shuvalov*, 2001, *Baldwin and Sheaffer*, 1971, *Borovicka et al.*, 1998]. For simplicity, henceforth we refer to all the material as meteoroids, whether they are intact from the moment they enter the atmosphere or result from the fragmentation of a larger body.

Surprisingly, the physico-chemical effects induced by the dispersed fragments in the upper atmosphere during the fragmentation episode have received little attention in literature. In particular, several important aspects of interaction of fragmented particles with the ambient atmosphere such as the local nitric oxide (NO) production and related ozone ($O_3$) destruction, especially during the non-ablative flight regime ($v \leq 10$ km/s), are poorly understood. This is further complicated if dispersed fragments are sufficiently large and still have necessary velocities to generate strong shockwaves [*Kadochnikov and Arsentiev*, 2018].

Comparatively, typical non-equilibrium hyperthermal chemistry, resulting from the interaction of the high temperature ablated meteor material with the local atmosphere, is reasonably well understood. Such processes that take place at extreme temperature (e.g. *Berezhnoy and Borovička*, 2010) produce large quantities of NO and other NOX products, destroy local ozone and have a capacity to strongly modify the ambient atmospheric region [e.g., *Menees and Park*, 1976, *Park and Menees*, 1978, *Berezhnoy and Borovička*, 2010, *Silber et al.*, 2017; *Silber et al.*, 2018a). Similarly, the chemistry of the thermalized meteor trails and their interaction with the ambient atmosphere have been studied extensively (e.g., *Baggaley*, 1978, *Baggaley*, 1979, *Plane*, 2012, *Plane et al.*, 2015 *Plane*, 2003, *Jenniskens*, 2004, *Whalley et al.*, 2011].

The production of NO in the high temperature meteor flow fields was recently investigated by [*Silber et al.*, 2018a], as NO plays a critical role in the structure and energetics of the upper atmosphere [*Marsh et al.*, 2004]. Silber et al. [*Silber et al.*, 2018a] considered only typical meteoric velocities and a limited subset of meteor sizes. However, NO is produced in all hypersonic flows exceeding a velocity of around 5 km/s (which is close to the dark flight regime) and its production is closely correlated to the degree of dissociation of $N_2$ and $O_2$. Below 95 km altitude, the mechanisms responsible for $N_2$ dissociation (e.g., soft X-rays, UV radiation and high energy auroral electrons) are typically less efficient in the thermosphere, and other sources of NO should be considered. Thus, the quantification of NO formed during the fragmentation of larger meteoroids may contribute toward resolving the uncertainties in the NO production in the region of mesosphere and lower thermosphere (MLT) where the natural production mechanisms are less efficient [*Silber et al.*, 2018a].

Here we report the results of the preliminary numerical study using ANSYS Fluent code to investigate the physico-chemical effect of cm-sized and larger fragmented objects in the non-ablative flight mode on the production of NO at altitudes 70 - 80 km. The model implemented in this work simulates a hypersonic chemically reactive flow over a spherical body (blunt body) in the continuum flow regime and accounts for the effects of dissociation and at higher velocity, the effect of ionization. It accurately simulates the production of NO within the boundaries of the flow field. However, our model is still limited in the sense that it cannot account for the intensive and non-linear nature of the meteoric ablation (which can be neglected at velocities below approximately 10 km/s) and the ensuing complex chemistry environment in high temperature meteor wakes (e.g., see [*Berezhnoy and Borovička*, 2010]). Thus, we approximate the typical meteor flow field by modeling only the atmospheric species hypersonic flow over a spherical body. For the purpose of this work, such approximation is valid as the production of NO does not appreciably depend on chemical reactions involving species from ablated high temperature meteoric plasma and vapour (e.g., [*Park and Menees*, 1978, *Berezhnoy and Borovička*, 2010]). The reasoning can be seen in the fact that the formation of meteoric metal monoxides takes place below 3,000 K, well below the peak optimal temperature for NO formation [*Berezhnoy and Borovička*, 2010]. Thus, the presence of a large number of chemical reactions involving ablated species in the meteor wake is assumed not to have a significant effect on NO production [*Menees and Park*, 1976, *Park and Menees*, 1978].

However, it should be noted that the actual flow field of a strongly ablating meteoroid at higher velocities is much larger than the characteristic object size (e.g., see [*Silber et al.*, 2018b, *Silber et al.*, 2018a]). Consequently, the fragment sizes were chosen to satisfy the flow field requirements corresponding to the continuous flow regime [*Silber et al.*, 2018b]. For the purpose of simplifying the computational approach in this exploratory study, we also neglect the compositional effect of the parent and fragmented objects and assume no appreciable size or mass loss in the studied velocity regime and the altitude interval. This study has important implications toward understanding the effect of intense meteor showers with many fragmentation events on the production of NO and a wider compositional and structural variability of the local atmosphere.

This paper is organized as follows; in Section 2 we describe the numerical approach, in Section 3 we present our results, discuss the outcomes in Section 4, and outline our conclusions in Section 5.

## 2. Numerical Methods

This section describes the numerical approach to investigate the production of NO by meteoroids. Three diameters ($d_m$) of solid spherical non-ablating objects were considered: $10^{-2}$, $10^{-1}$ and 1 m, at altitudes ($h$) of 70 km and 80 km. We performed simulations for two velocities ($v$), 5 km/s and 10 km/s. The sizes represent the extra-terrestrial objects capable of generating a shock wave given appropriate conditions (see the Introduction section). Additionally, the chosen sizes are within the group of the smallest shock-forming meteoroids that impact the Earth's atmosphere, and incidentally are also the most frequent [*Drolshagen et al.*, 2017, *Silber et al.*, 2017]. Moreover, the dimensions given are consistent with the size of fragments that would result from a disruption episode during a meteoroid flight through the atmosphere. The choice for the velocities is two-fold. First, the fragments released as a result of a disruption episode will tend to slow down faster than a more massive parent body. Second, the numerical models best describe the motion and physical and chemical behaviour of hypersonic objects at lower velocities. For a detailed discussion about the challenges in numerically modeling meteoroids, the reader is referred to [*Silber et al.*, 2018b].

The axisymmetric 2D numerical simulations were performed using ANSYS Fluent code version 18.0, a proprietary computational fluid dynamics (CFD) Eulerian code based on the finite volume method (www.ansys.com). ANSYS Fluent includes the AUSM+ scheme [*Liou*, 2006] for the discretization of convective fluxes, suitable for hypersonic flow. The code is suitable for simulating supersonic and hypersonic flows because it can solve the Navier-Stokes equations written in the conservative form, which minimizes the numerical errors when discontinuities (shock waves) appear. Furthermore, it has been used extensively in similar studies [*Niculescu et al.*, 2016; *Silber et al.*, 2017; *Niculescu et al.*, 2018]. Since ANSYS is proprietary, the inner workings of this code will be omitted here. The governing equations for hypersonic flow are given in [*Niculescu et al.*, 2016]. This code is not optimized for the inclusion of ablation, radiation and excitation, but it does account for ionization and it uses the non-equilibrium approach. The chemical reactions, including dissociation of $N_2$ and $O_2$, and formation of NO and ionized species are included in the code.

The convective flux was discretized with AUSM+-up scheme [*Liou*, 2006] because this scheme has been proven to work well for hypersonic gas dynamics [*Kitamura*, 2016, *Kitamura and Shima*, 2013]. It is worth to mention that while other alternative and interesting schemes and solvers exist (e.g., [*Zeidan et al.*, 2019; *Goncalvès and Zeidan*, 2017; *Zeidan and Touma*, 2014; *Zeidan et al.*, 2007]), those are not feasible to implement due to the proprietary nature of the ANSYS code, and as such have not been tested for this specific application. Due to its robustness, we have preferred the implicit formulation rather than the explicit one. More details regarding the CFD methodology is given in [*Niculescu et al.*, 2016]. The classic governing equations for hypersonic gas dynamics in thermal equilibrium are the Navier-Stokes equations (see Appendix A and [*Niculescu et al.*, 2016]). However, it is necessary to also add the transport equations of species that appear/disappear due to the chemical reactions. The detailed outline of this can be found in [*Chung*, 2010; *Viviani and Pezzella*, 2015].

At the external boundary of computational domain, all flow parameters were imposed, i.e., the velocity of meteoroid (5 km and 10 km), pressure and temperature of air corresponding to altitudes of 70 km and 80 km and chemical composition of air (21% $O_2$ and 79% $N_2$). At the outlet boundary, all flow parameters are extrapolated from adjacent cells according to theory of characteristics for supersonic flows. The mesh is fully structured and multiblock. It contains about 200,000 quadrilateral cells clustered near the meteor and the minimal height of cells adjacent to wall is $10^{-5}$ m.

At Mach numbers up to 20, the temperature field is too low to allow a significant ionization [*Niculescu et al.*, 2019 and references therein]. For this reason, it is preferable to use a chemical model that takes into account only dissociation of oxygen and nitrogen and formation/destruction of NO such as the Park'89 model [*Park*, 1989; *Viviani and Pezzella*, 2015] in order to appreciably decrease the computational effort and complexity of chemical model, which is prone to numerical errors. At speeds of 10 km/s, the Mach number is about 35. In this case, all air is dissociated and the ionization is significant; therefore, it should be included in the chemical model. For this reason, the Dunn and Kang model [*Dunn and Kang*, 1973; *Anderson*, 2000] was implemented in ANSYS Fluent. The ionization greatly influences the temperature field because it is an endothermic phenomenon; therefore, it should be taken into account. Tables with chemical reactions for both models are given in Appendix B.

The computational domain extends 280 times the meteoroid diameter in the axial direction behind the object, and 80 times in the radial direction. We obtained the pressure, temperature, mol concentration of NO and mass fraction of NO for all our simulations in order to investigate their dependence on meteoroid size and velocity.

## 3. Results

Our results show that the effects of the meteoroid velocity and altitude on shock wave formation are as expected. A meteoroid traveling at higher velocity will generate a stronger shock wave which subsequently produces higher temperature flow fields and are conducive to higher rates of NO production. Higher air density at lower altitudes will also promote NO production. Here we explore how these quantities differ as a function of meteoroid size, velocity and height for the scenarios used in our simulations.

The meteoroid velocity plays a significant role; however, as seen in Figure 1, its effect on the production of NO strongly depends on the object diameter (i.e., the size of the initial flow field envelope). In Figure 1, plotted are the ratio of the cumulative mass ($m_{NO}$) of NO produced at 10 km/s and 5 km/s (i.e., $m_{NO\_10km/s}/m_{NO\_5km/s}$) versus radial distance from the meteoroid propagation axis. The greatest velocity difference can be seen for the 1 cm meteoroid – there is about 30 times more NO produced at 10 km/s versus 5 km/s velocity. However, for the 10 cm meteoroid, this difference is only a factor of ~7, and for a 1 m object, the difference is negligible.

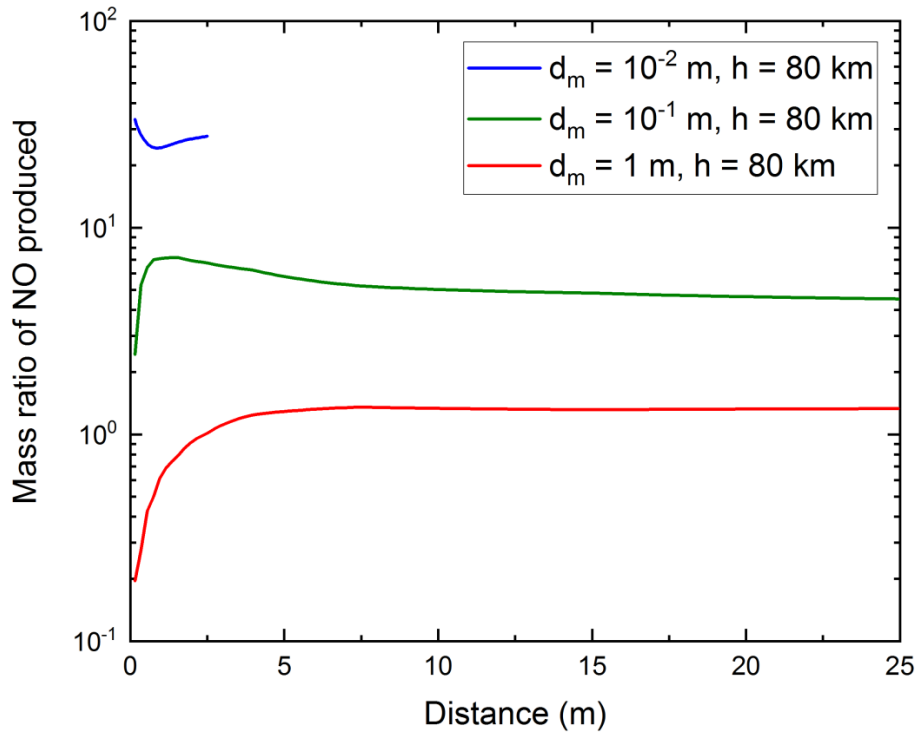

**Figure 1:** The ratio of cumulative mass of NO produced by meteoroids travelling at 80 km altitude at 10 km/s and 5 km/s. The vertical axis represents the velocity ratio $m_{NO\_10km/s}/m_{NO\_5km/s}$.

Figure 2 shows the NO production dependence on the altitude while keeping the velocity constant ($v$ = 10 km/s). Plotted is the ratio of the cumulative mass ($m_{NO}$) of NO produced at 70 km and 80 km altitude (i.e., $m_{NO\_70km}/m_{NO\_80km}$) as a function of radial distance from the meteoroid propagation axis. The two smallest meteoroids exhibit a different behaviour relative to the 1m object, most notably up to the distance of 5 m from the propagation axis. There is an apparent dip in the NO mass ratio for the two smallest meteoroids, and a spike in the NO production for a 1m object. The apparent difference in the behaviour of the 1m object in both Figure 1 and Figure 2 is because for large objects ($d_m > 1$ m), the recompression shock wave becomes an important source of NO at speeds greater than ~5 km/s (the Mach number is ~17) [*Silber et al.*, 2018a]. Moreover, if the meteoroid generated NO undergoes diffusion and enters the flow field streamline with a temperature higher than 6,000 K, it will be destroyed relatively quickly.

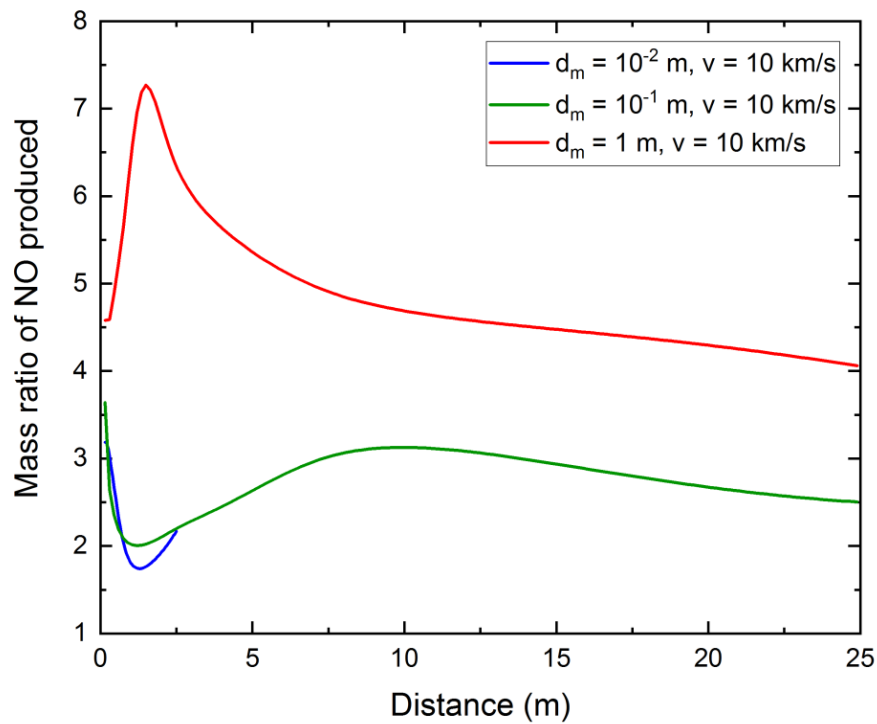

**Figure 2:** The ratio of cumulative mass of NO produced by meteoroids travelling at 10 km/s for altitudes 70 and 80 km. The vertical axis represents the ratio $m_{NO\_70km} / m_{NO\_80km}$.

The mass fraction of NO and the temperature field produced by a meteoroid with the diameter of $10^{-2}$ m are plotted in Figure 3. The panels on the left show the mass fraction of NO, and the panels on the right represent the temperature of the flow field at various altitudes and meteoroid speeds. These correspond to, as seen from top to bottom, $v = 5$ km/s, $h = 80$ km (Figure 3a,b), $v = 10$ km/s, $h = 70$ km (Figure 3c,d), and $v = 10$ km/s, $h = 80$ km (Figure 3e,f). We note that the colour bar scale is not the same for all panels and this is done intentionally, in order to highlight the differences in peak mass fraction of NO, as well as the peak temperature for each scenario.

For comparison, we also plot the same quantities for a $10^{-1}$ m meteoroid in Figure 4. The initial production of NO around the meteoroid and immediately behind it (within the shock envelope boundary) is mainly concentrated on the outer periphery of the flow field. Comparatively, both the velocity and size have an effect on the spatially very small initial region of the NO equilibrium flow, which is in line with the discussion presented by [*Berezhnoy and Borovička*, 2010] for bodies at 80 km altitude.

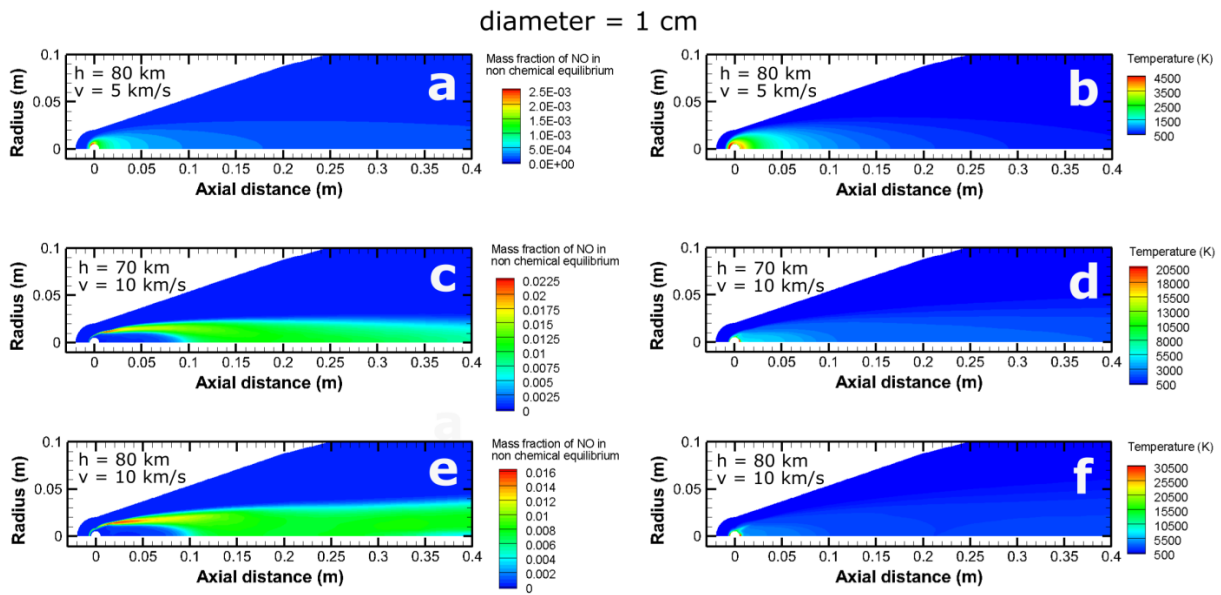

**Figure 3:** Mass fraction of NO (panels on the left) and temperature field (panels on the right) for a 1 cm meteoroid traveling at (a,b) 5 km/s at 80 km altitude, (c,d) 10 km/s at 70 km altitude and (e,f) 10 km/s at 80 km altitude. Note that the colour bar is not scaled the same across all panels.

The spatial distribution of NO production is as expected, since only $N_2$ and $O_2$ that survive the initial collision with the shock envelope impact the outer region ending up in streamlines of the "frozen" flow (significantly reduced number of collisions) (e.g., [*Anderson*, 2000]). These streamlines eventually merge with the main flow field in the turbulent region further in the wake and provide a supply of reactants to chemical reactions, promoting the production of NO within the flow field (Figure 4a,c,d and Figure 4a,c,d). Both high velocity and larger size are correlated with higher peak temperatures. The falloff in the temperature is more rapid in smaller and slower meteoroids. The peak temperatures along the axis show values similar to those obtained by [*Boyd*, 2000].

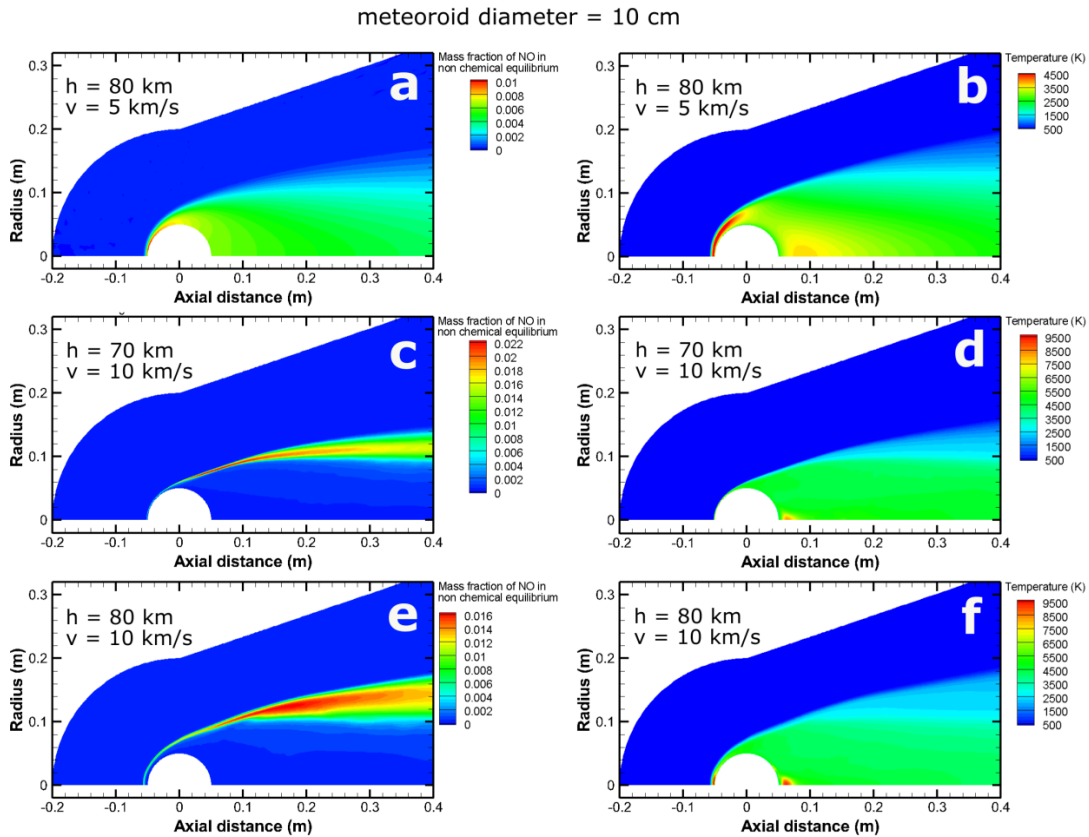

**Figure 4:** Mass fraction of NO (panels on the left) and temperature field (panels on the right) for a 10 cm meteoroid traveling at (a,b) 5 km/s at 80 km altitude, (c,d) 10 km/s at 70 km altitude and (e,f) 10 km/s at 80 km altitude. Note that the colour bar is not scaled the same across all panels.

The cumulative mass of NO produced within the boundaries of the modeled flow field for all our simulations are shown in Figure 5. The length of the plotted interval corresponds to the size of the computational domain. Therefore, the lines associated with the NO production for $1 \cdot 10^{-2}$ m meteoroid are limited to 2.5 m because this is the length of the computational domain for this meteoroid size. In Figure 6, we show the NO production for 1 cm meteoroid in panel (a) and 10 cm meteoroid in panel (b) for better visualization. The rate of NO production decreases away from the optimal temperature region and approaches the finite value in the region beyond ~15 m away from the meteoroid. Essentially, the production of NO continues as long as the temperature is higher than approximately 2,000 K. Beyond ~15 m, the shock wave is too weak to generate high enough temperature that would be conducive to the production of NO. This behaviour is in line with the interpretation given for large meteoroids in [*Silber et al.*, 2018a].

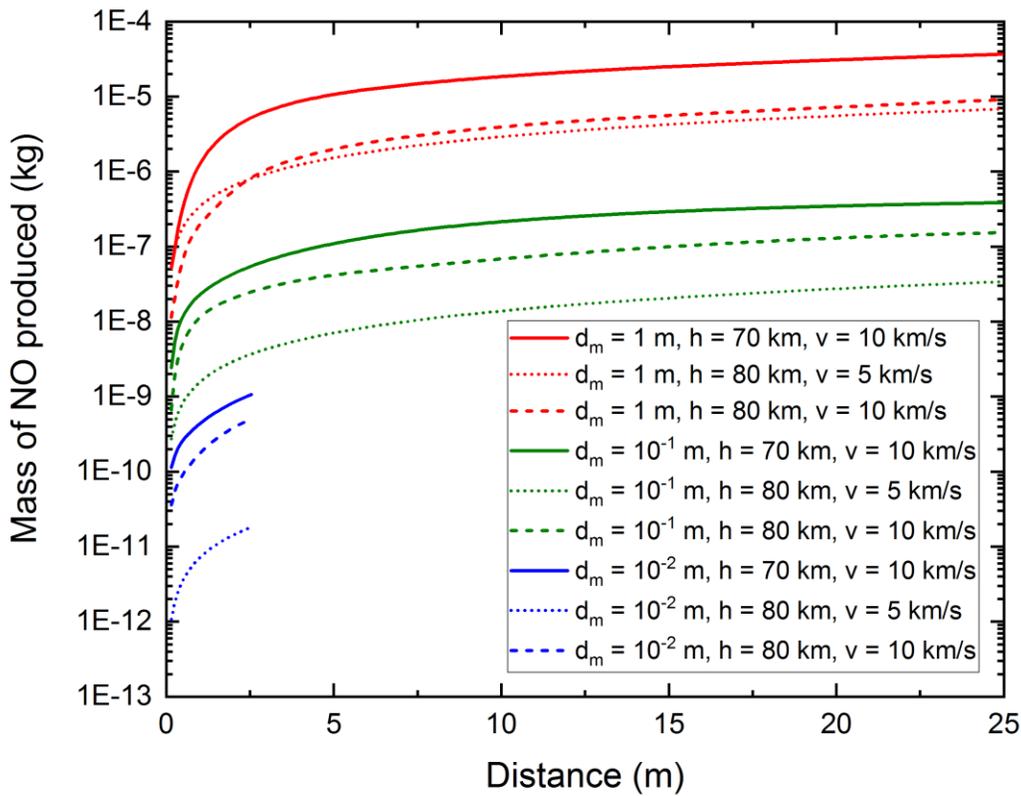

**Figure 5:** The cumulative mass of NO produced within the boundaries of the modeled flow field for all our simulations.

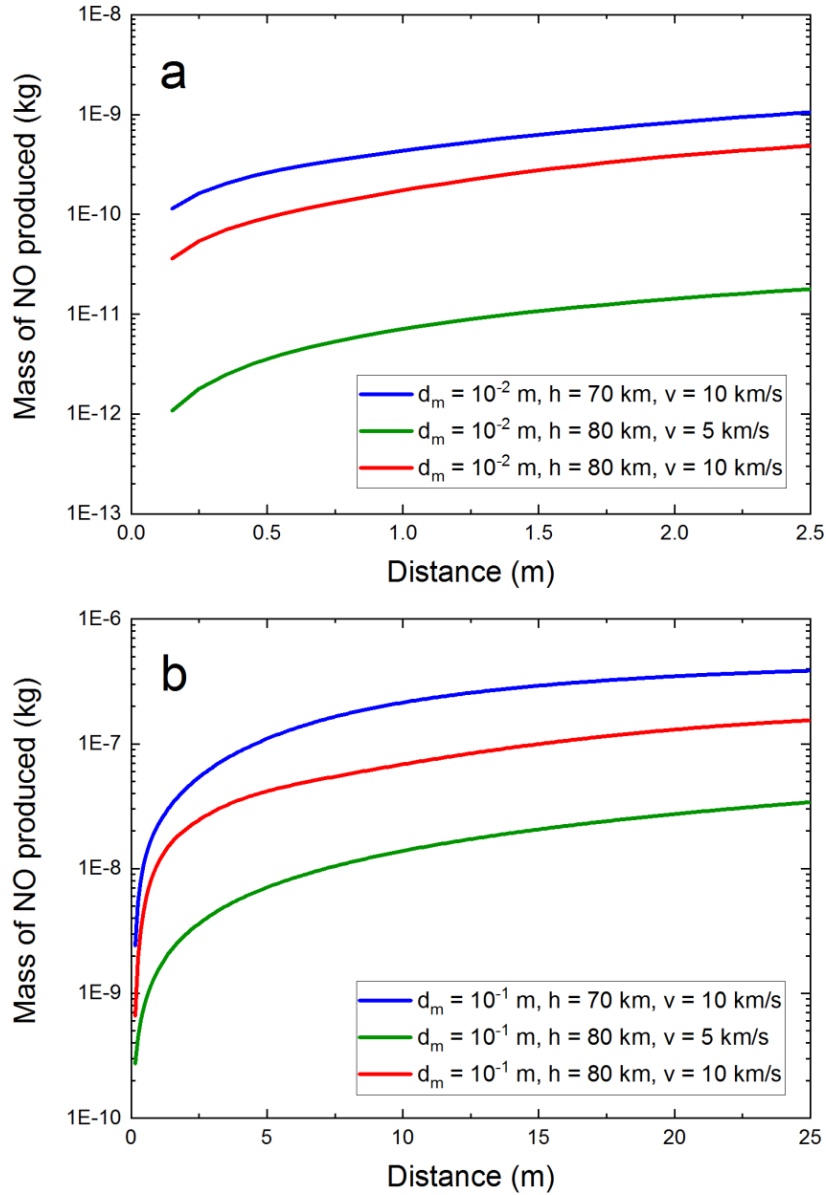

**Figure 6:** The cumulative mass of NO produced within the boundaries of the modeled flow field for (a) 1 cm and (b) 10 cm meteoroids.

Figures 7 – 9 show the radial temperature profiles for all our simulations for a number of distance points behind the meteoroid. Again, as expected, it is the size and velocity that play the dominant role in the temperature distribution in the flow field. The profiles for a $1 \cdot 10^{-2}$ m meteoroid (Figure 7) travelling at 5 km/s at 80 km altitude demonstrate that the temperature is too low to promote reactions required to produce any appreciable amount of NO. This links back to our initial statement related to Figure 1 earlier in this section.

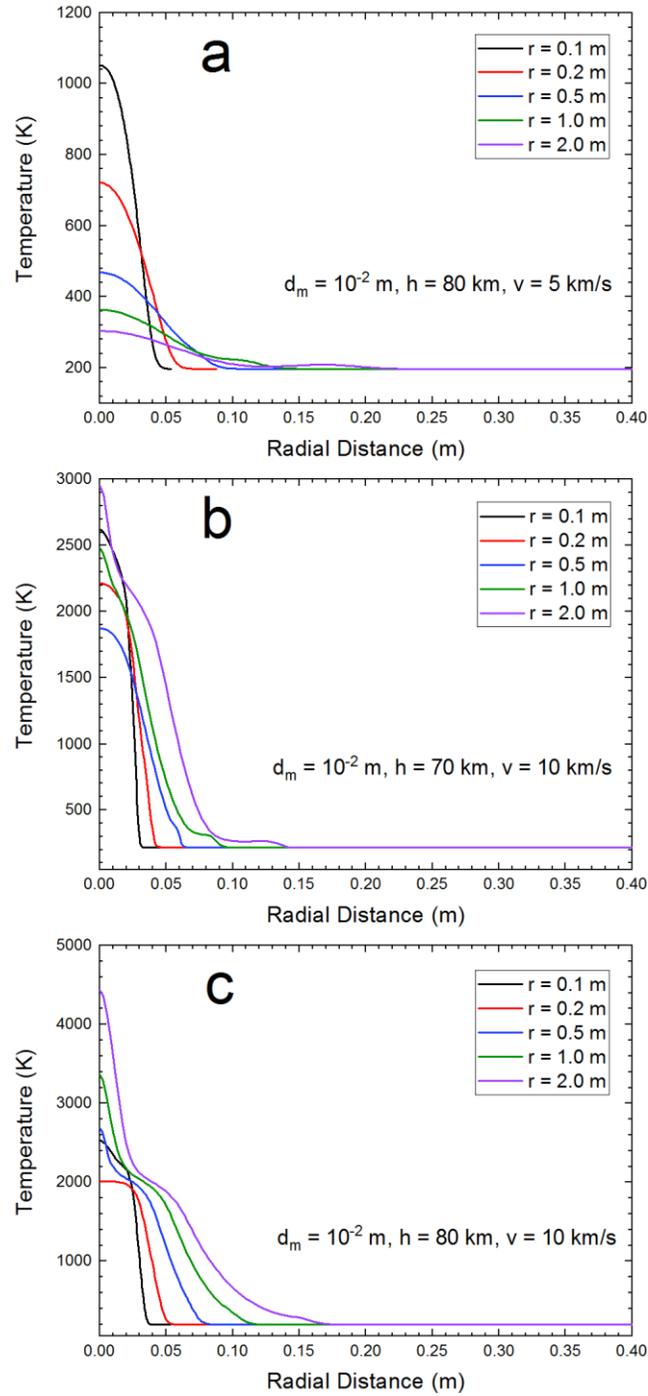

**Figure 7:** The radial temperature profiles at various distances for a meteoroid with $d_m = 10^{-2}$ m: (a) $h = 80$ km, $v = 5$ km/s (b) $h = 70$ km, $v = 10$ km/s, (c) $h = 80$ km, $v = 10$ km/s. Note that the vertical scale varies across the panels.

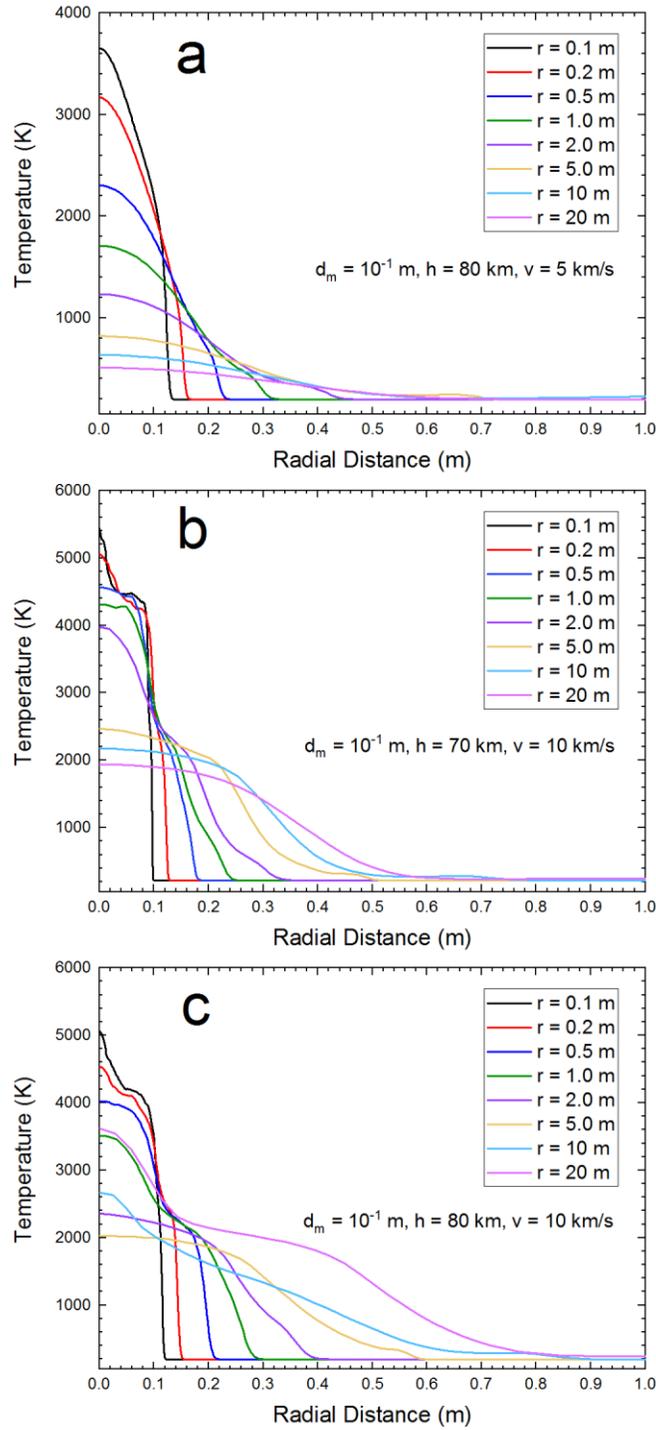

**Figure 8:** The radial temperature profiles at various distances for a meteoroid with $d_m = 10^{-1}$ m: (a) $h = 80$ km, $v = 5$ km/s (b) $h = 70$ km, $v = 10$ km/s, (c) $h = 80$ km, $v = 10$ km/s. Note that the vertical scale varies across the panels.

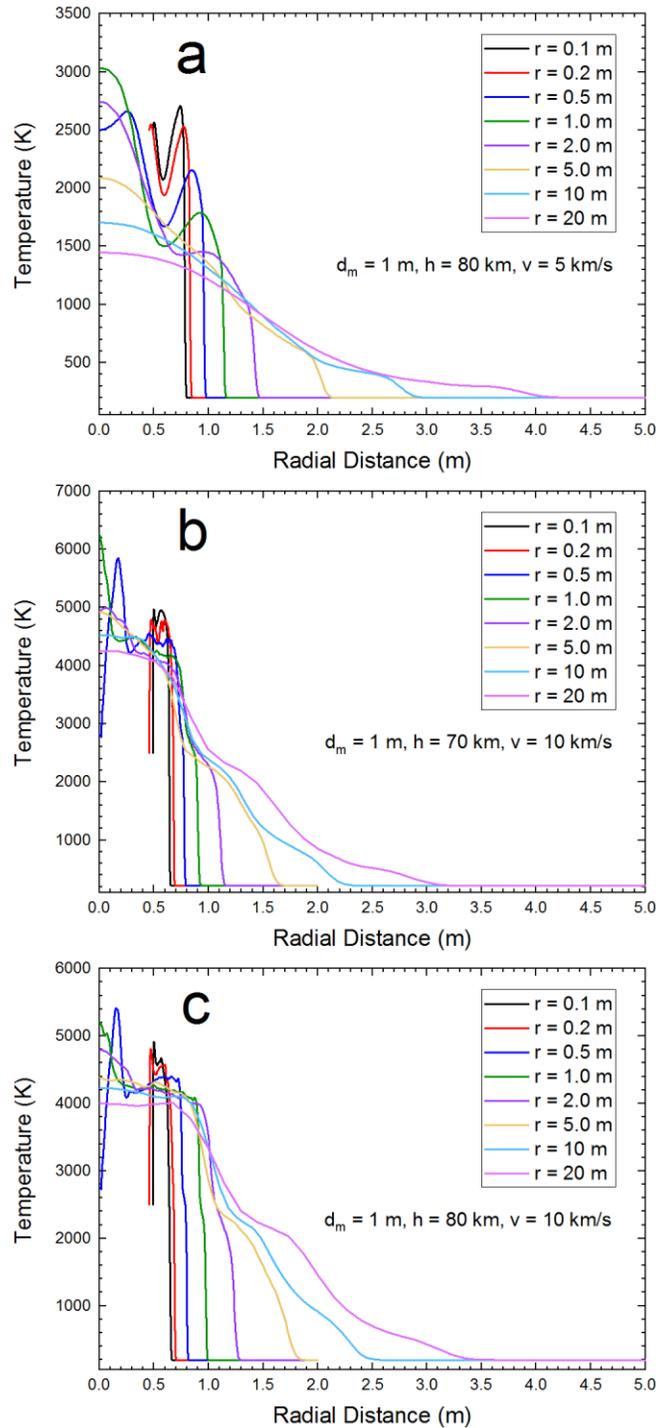

**Figure 9:** The radial temperature profiles at various distances for a meteoroid with $d_m = 1$ m: (a) $h = 80$ km, $v = 5$ km/s (b) $h = 70$ km, $v = 10$ km/s, (c) $h = 80$ km, $v = 10$ km/s. Note that the vertical scale varies across the panels.

Using the number flux rate for centimeter-sized meteoroids published by [*Drolshagen et al.*, 2017], assuming a very small rate of fragmentation (corresponding to the flux of impactors) and then extrapolating the NO production from our simulations for the layer of the atmosphere between 70 and 80 km altitude, we estimate the lowermost limit of the annual NO production at 3.5 tons/year. However, this value could be an order of magnitude smaller than the actual value, because it applies to the limited, albeit most representative, region of the atmosphere (70 – 80 km) and does not take into account all fragmentation events, as such has not been studied in any appreciable detail. Further discussion is given in the next section.

## 4. Discussion

The preliminary results obtained in this study are consistent with expected values for shock producing bodies at MLT altitudes (c.f. [*Sarma*, 2000] and references therein). Increasing velocities result in a higher initial temperature in the flow field (Figures 1 and 2) and subsequently lead to an increase of NO production. However, higher velocities also impact the ratio of hydrodynamic and chemistry time scales [e.g., *Berezhnoy and Borovička*, 2010], which subsequently shift the peak production of NO slightly further away from the aft of the body along the axis of propagation. The rate of NO production approaches the finite value in the region beyond ~15 m away from meteoroid (for an object with the diameter of $10^{-1}$ and 1 m). This is consistent with the reasoning that the highest NO production takes place during the velocity dependent meteor radiative stage (up to several hundred meters at MLT altitudes for cm-sized objects) [e.g., *Jenniskens and Stenbaek-Nielsen*, 2004] and has been observed in an earlier study [*Menees and Park*, 1976]. Of course, the production of NO slows down significantly and eventually ceases when the flow becomes significantly rarefied, even if higher temperatures are present. This can be seen from Figures 1 and 2.

Larger and faster events produce more NO within the boundaries of the flow field. This is expected as the size of the flow field is larger, and these events have higher temperatures persisting longer distances in the flow field behind the body. The point to emphasize here is that the initial production of NO around the meteoroid and immediately behind it (within the shock envelope boundary) is mainly concentrated on the outer periphery of the flow field. Comparatively, both the velocity and size have an effect on the spatially very small initial region of the NO equilibrium flow, which is in line with the discussion presented by [*Berezhnoy and Borovička*, 2010] for bodies at 80 km altitude. The apparent difference in the behaviour of the NO mass ratio as a function of velocity and altitude for a 1 m object (see Figures 1 and 2) is due to the fact that larger objects (as opposed to cm-sized meteoroids) exhibit a more prominent recompression shock, which also pays a role in the NO production. This is consistent with the interpretation given by [*Silber et al.*, 2018a].

Our lowermost estimate for NO produced by centimetre-sized meteoroids is approximately 3.5 tons/year. However, this value could be an order of magnitude underestimated, because we assumed the process to occur only in a narrow region of the atmosphere ($70 - 80$ km) with limited fragmentation. It should be noted that constraining the altitude to $70 - 80$ km is still a reasonable assumption because most small meteoroids burn up in the MLT due to ablation, and thus never penetrate deeper into the atmosphere [e.g., *Silber and Brown*, 2014, *Silber et al.*, 2018b and references therein]. However, the rate of fragmentation for centimetre-sized objects is difficult to constrain because statistical studies on the topic are absent. Studies concerning meteorite delivery to Earth suggest that most meteoroids undergo some sort of fragmentation (either continuous or gross fragmentation) [e.g., *Ceplecha et al.*, 1998]. Nevertheless, our results are comparable to the recently published estimate [*Silber et al.*, 2018a]. This reaffirms the notion that while centimetre-sized meteoroids are efficient producers of NO, their cumulative contribution could be relatively negligible.

The biggest uncertainties in NO production in this study are closely related to the meteoroid fluxes with different sizes, velocities and fragmentation rates. While we conservatively assumed that the fragmentation rate corresponds to the meteoroid fluxes in the given size regime, this might be a gross underestimate. Therefore, our lowest limit of NO production by the range of meteor fragment sizes may admittedly be underestimated by as much as an order of magnitude. Indeed, this will need to be clarified in future studies as the meteoroid flux data and fragmentation rates become more refined. Moreover, it is clear from our study that the small and rapidly decelerating fragments will contribute a negligible amount to the overall NO production. Additionally, as the species production in reactive meteoroid flow fields is governed by the Damköhler number, it also means that the ambient diffusion plays a very important role in NO production as hinted by [*Hocking et al.*, 2016].

It is worth to recall that the two regimes of NO production, namely cold and hot regime, which are strongly dependent meteoroid fragment size and velocity interval (for a detailed discussion see [*Silber et al.*, 2018a]). The existence of both regimes is also related to the Damköhler number and may be preferentially dominant in different size and temperature flow fields. The first regime is temperature governed and it is associated with larger and hyperthermal flow fields, while the second regime depends on $O_2$ availability and is preferentially present in rarefied flow fields generated by smaller events. This regime is also comparably inefficient, which is in line with previous work (see [*Silber et al.*, 2018a] for detail). However, in the case of large meteoroids, the initial shock is very strong and capable of dissociating most of $O_2$ and $N_2$ in the wider region of the ambient atmosphere surrounding the axis of the meteoroid propagation (*Silber et al.*, 2018a). In this case, the supply of $N_2$ and $O_2$ becomes very limited with increasing meteoroid particle size, thereby reducing the efficiency of NO production.

Smaller fragments are incapable of generating strong shockwaves necessary for the dissociation of atmospheric species that contribute to NO production (e.g., [*Silber et al.*, 2018a]). That is primarily because the flow filed associated with small fragments rapidly rarefies and almost instantaneously thermalizes behind the particle [*Silber et al.*, 2017]. Thus, as mentioned earlier, small and rapidly decelerating fragments will have a negligible contribution to the overall NO production. On the other hand, the removal of NO from hypersonic high temperature rarefied flows remains poorly constrained due to the lack of comprehensive experimental data at high altitudes.

We also wish to add a relevant point about numerically modelling meteoroids and meteoroid fragments. Comprehensive modeling of meteor flows in a rarefied environment of MLT is extremely difficult as classical Navier-Stokes equations are not valid in such conditions. A suitable model would need to account for the strong ablation while accounting for nonlinear heat transfer to the body (e.g., [*Wang*, 2014]). Consequently, such a simulation would also need to account for typical shock layer related effects such as vibrational excitation, dissociation, electronic excitation, ionization and radiation phenomena in the rarefied gas. All of these effects associated with the rarefied and hypersonic flows are a consequence of the non-equilibrium real gas effects and are nonlinear, which implies that their simplifications can lead to significant errors in the model predictions.

An additional challenging and difficult modeling task in the near wake of the meteoroid is simulating the flow of ablated vapour and plasma that starts rapidly expanding as it emerges from the "recompression" region. This flow forms a high temperature turbulent diffusive region which is highly nonlinear (fundamentally, a mean state of spatiotemporal chaos) that eventually leads to the formation of the dynamically stable volume of meteor plasma with some initial radius. The importance of that region is in the fact that it is responsible for mixing of ablated and entrained species that participate in the temperature driven chemistry. The computational approach should be able to include a detailed description of such a non-linear dynamical system and to account for all physico-chemical phenomena. To date, no such comprehensive model exists that accounts for all the effects associated with the rarefied hypersonic flow of a typical meteoroid. For a detailed discussion on intricacies pertaining to meteor generated shock waves, see the review by [*Silber et al.*, 2018b].

## 5. Conclusions

All optically detectable meteors, as well as many of the stronger radio-detectable meteors, produce shockwaves and fragment during the lower transitional and continuum flow regimes in the MLT region of the atmosphere (*Silber et al.*, 2018b, *Silber et al.*, 2017). Meteoroid-fragment generated shock waves can modify the surrounding atmosphere and produce a range of physico-chemical effects. Some of the thermally driven chemical and physical processes induced by meteoroid fragment generated shock waves, such as NO production, are less understood. Any estimates of meteoric fragment NO production depend not only on a quantifiable sufficiently large meteoroid population with a size capable of producing fragments generating high temperature flows, but also on understanding the physical properties of the meteor flows along those fragments with their thermal history.

We performed an exploratory pilot study to investigate the production of NO by meteoroids (or fragments of meteoroids after they undergo a disruption episode), in the size range from $10^{-2}$ to 1 m (diameter) in the MLT region of the atmosphere. Our model uses the simulation of a spherical body in the continuum flow at 70 and 80 km altitude to approximate the behaviour of a small meteoroid capable of producing NO. The minimal annual NO production by meteoroid fragmentation is estimated at about 3.5 tons/year. However, this estimate may be understated by as much as an order of magnitude, considering the large uncertainties in fluxes of meteoroids with specific size and velocities as well large uncertainties in fragmentation rates. Nevertheless, the initial results presented in this exploratory study are in good agreement with previous studies.

**Acknowledgements**

The authors thank the two anonymous reviewers for their thoughtful comments that helped improve this paper.

## Appendix A

Here we list the classic governing equations for hypersonic gas dynamics in thermal equilibrium (also see [*Niculescu et al.*, 2016]).

$$\frac{\partial yU}{\partial t} + \frac{\partial y(F_x - G_x)}{\partial x} + \frac{\partial y(F_y - G_y)}{\partial y} = S$$

where $U = \begin{pmatrix} \rho \\ \rho u \\ \rho v \\ \rho E \\ \rho Y_i \end{pmatrix}$ − conservative variables; $Y_i$ − mass fraction of species $i$

$$F_x = \begin{pmatrix} \rho u \\ \rho u^2 + p \\ \rho uv \\ \rho uH \\ \rho uY_i \end{pmatrix}, \ F_y = \begin{pmatrix} \rho v \\ \rho vu \\ \rho v^2 + p \\ \rho vH \\ \rho vY_i \end{pmatrix} - \text{convective fluxes}$$

$$G_x = \begin{pmatrix} 0 \\ \tau_{xx} \\ \tau_{xy} \\ u\tau_{xx} + v\tau_{xy} + k\frac{\partial T}{\partial x} + \rho \sum_{i=1}^{N} h_i D_{i\,mixture} \frac{\partial Y_i}{\partial x} \\ \rho D_{i\,mixture} \frac{\partial Y_i}{\partial x} \end{pmatrix} - \text{diffusive flux in axial direction}$$

$$G_y = \begin{pmatrix} 0 \\ \tau_{yx} \\ \tau_{yy} \\ u\tau_{yx} + v\tau_{yy} + k\frac{\partial T}{\partial y} + \rho \sum_{i=1}^{N} h_i D_{i\,mixture} \frac{\partial Y_i}{\partial y} \\ \rho D_{i\,mixture} \frac{\partial Y_i}{\partial y} \end{pmatrix} - \text{diffusive flux in radial direction}$$

$$S = \begin{pmatrix} 0 \\ 0 \\ p - \tau_{\theta\theta} \\ -y \sum_{i=1}^{N} \left[ h_i^{formation\,T_{ref}} - c_{p_i}(T_{ref})T_{ref} \right] \omega_i \\ y\omega_i \end{pmatrix} - \text{source term}$$

$\omega_i$ – rate of formation/destruction of species $i$, $N$ – number of chemical species,

$\tau_{\theta\theta} = \frac{2}{3}\mu\left(2\frac{v}{y} - \frac{\partial u}{\partial x} - \frac{\partial v}{\partial y}\right)$ – cross viscous stress

$\tau_{xx} = \frac{2}{3}\mu\left(2\frac{\partial u}{\partial x} - \frac{\partial v}{\partial y} - \frac{v}{y}\right)$ – axial viscous stress

$\tau_{yy} = \frac{2}{3}\mu\left(2\frac{\partial v}{\partial y} - \frac{\partial u}{\partial x} - \frac{v}{y}\right)$ – $r$adial viscous stress

$\tau_{xy} = \tau_{yx} = \mu\left(\frac{\partial u}{\partial y} + \frac{\partial v}{\partial x}\right)$ – transversal viscous stress (the same as for 2D planar flows)

$p = \rho R^0 T \sum_{i=1}^{N} \frac{Y_i}{W_i}$ – equation of idel gas , $R^0 = 8314.3 \frac{J}{kmol.K}$

$W_i$ – molecular weight of species i

$h_i = h_i(T) = c_{p_i}(T)T$ – hypothesis of a calorically perfect gas for chemical species

$\mu_i(T) = \mu_{0_i}\left(\frac{T}{T_{0_i}}\right)^{1.5} \frac{T_{0_i} + S_i}{T + S_i}$ – Sutherland law with 3 coefficients

$$\sum_{i=1}^{N} \nu'_{ij} M_i \xrightarrow{k_f} \sum_{i=1}^{N} \nu''_{ij} M_i$$

$j$ - number of chemical reactions

$\omega_i = W_i \sum_{j=1}^{number\ of\ reactions} \left[(\nu''_{ij} - \nu'_{ij})\Gamma_j k_{f_j} \prod_{i=1}^{N}\left(\frac{X_i p}{R^0 T}\right)^{\nu'_{ij}}\right]$

$X_i = Y_i \frac{W}{W_i}$, $\frac{1}{W} = \sum_{i=1}^{N}\frac{Y_i}{W_i}$, $k_{f_j} = A_j T^{\beta_j} e^{-\frac{E_{a_j}}{R^0 T}}$ – $Arrhenius\ law$

$X_i$ – volume (molar) fraction of species $i$, $W$ – molecular weight of mixture

$k_{fj}$ -rate of forward chemical reaction $j$, $E_j$ – activation energy of reaction $j$

**Appendix B**

**Table B1**: The model of chemical reactions of dissociation and ionization of air proposed by [*Dunn and Kang*, 1973, *Anderson*, 2000], which is suitable for velocity of meteoroid propagating at velocities of 10 km/s and greater.

| Reaction | $A_f$ [$K^{-\beta_f}/s$] | $\beta_f$ | $E_{af}$ [$J/kmol$] |
|---|---|---|---|
| $O_2 + N = 2O + N$ | 3.6000E15 | -1.00000 | 4.97E+08 |
| $O_2 + NO = 2O + NO$ | 3.6000E15 | -1.00000 | 4.97E+08 |
| $N_2 + O = 2N + O$ | 1.9000E14 | -0.50000 | 9.45E+08 |
| $N_2 + NO = 2N + NO$ | 1.9000E14 | -0.50000 | 9.45E+08 |
| $N_2 + O_2 = 2N + O_2$ | 1.9000E14 | -0.50000 | 9.45E+08 |
| $NO + O_2 = N + O + O_2$ | 3.9000E17 | -1.5 | 6.31E+08 |
| $NO + N_2 = N + O + N_2$ | 3.9000E17 | -1.5 | 6.31E+08 |
| $O + NO = N + O_2$ | 3.2000E06 | 1 | 1.65E+08 |
| $O + N_2 = N + NO$ | 7.0000E10 | 0 | 3.18E+08 |
| $N + N_2 = 2N + N$ | 4.0850E19 | -1.5 | 9.45E+08 |
| $O + N = NO^+ + e^-$ | 1.4000E03 | 1.50000 | 2.67E+08 |
| $O + e^- = O^+ + 2e^-$ | 3.6000E28 | -2.91 | 1.32E+09 |
| $N + e^- = N^+ + 2e^-$ | 1.1000E29 | -3.14 | 1.41E+09 |
| $O + O = O_2^+ + e^-$ | 1.6000E14 | -0.98000 | 6.75E+08 |
| $O + O_2^+ = O_2 + O^+$ | 2.9200E15 | -1.11000 | 2.34E+08 |
| $N_2 + N^+ = N + N_2^+$ | 2.0200E8 | 0.81000 | 1.09E+08 |
| $N + N = N_2^+ + e^-$ | 1.4000E10 | 0 | 5.67E+08 |
| $O + NO^+ = NO + O^+$ | 3.6300E12 | -0.6 | 4.25E+08 |
| $N_2 + O^+ = O + N_2^+$ | 3.4000E16 | -2.00000 | 1.92E+08 |
| $N + NO^+ = NO + N^+$ | 1.0000E16 | -0.93 | 5.10E+08 |
| $O_2 + NO^+ = NO + O_2^+$ | 1.8000E12 | 0.17000 | 2.76E+08 |
| $O + NO^+ = O_2 + N^+$ | 1.3400E10 | 0.31 | 6.46E+08 |
| $O_2 + O = 2O + O$ | 9.0000E16 | -1 | 4.97E+08 |
| $O_2 + O_2 = 2O + O_2$ | 3.2400E16 | -1 | 4.97E+08 |
| $O_2 + N_2 = 2O + N_2$ | 7.2000E15 | -1 | 4.97E+08 |
| $N_2 + N_2 = 2N + N_2$ | 4.7000E14 | -0.5 | 9.45E+08 |
| $NO + O = N + 2O$ | 7.8000E17 | -1.5 | 6.31E+08 |
| $NO + N = O + 2N$ | 7.8000E17 | -1.5 | 6.31E+08 |
| $NO + NO = N + O + NO$ | 7.8000E17 | -1.5 | 6.31E+08 |
| $O_2 + N_2 = NO + NO^+ + e^-$ | 1.3800E17 | -1.84 | 1.18E+09 |
| $NO + N_2 = NO^- + e^- + N_2$ | 2.2000E12 | -0.35 | 9.03E+08 |

**Table B2:** The reaction rate parameters in Park'89 model [*Park*, 1989, *Viviani and Pezzella*, 2015], which is suitable for velocity of meteoroid travelling at 5 km/s or less because the temperature field does not allow a significant ionization [*Niculescu et al.*, 2019].

| No | Reaction | $A_{f,r}$ $[K^{-\beta}/s]$ | $\beta_{f,r}$ | $E_{a\ f,r}$ $[J/kmol]$ | Third body efficiency $\Gamma$ | Heat of reaction at 298.15 K $[kJ/mol]$ |
|----|----------|------|------|------|------|------|
| 1 | $O_2+M \rightarrow 2O+M$ | 1.00E19 | -1.5 | 4.947E8 | $O_2=N_2=NO=0.2$, $O=N=1$ | 498.56 |
| 2 | $N_2+M \rightarrow 2N+M$ | 3.00E19 | -1.6 | 9.412E8 | $O_2=N_2=NO=0.233$, $O=N=1$ | 945.92 |
| 3 | $NO+M \rightarrow N+O+M$ | 1.10E14 | 0.0 | 6.277E8 | $O_2=N_2=0.05$, $O=N=NO=1$ | 631.92 |
| 4 | $NO+O \rightarrow O_2+N$ | 2.40E6 | 1.0 | 1.598E8 | - | 133.36 |
| 5 | $N_2+O \rightarrow NO+N$ | 1.80E11 | 0.0 | 3.193E8 | - | 314 |